# A Rare Topic Discovery Model for Short Texts Based on Co-occurrence word Network

Chengjie Ma, Junping Du, Yingxia Shao, Ang Li, Zeli Guan

**Abstract.** We provide a simple and general solution for the discovery of scarce topics in unbalanced short-text datasets, namely, a word co-occurrence network-based model CWIBTD, which can simultaneously address the sparsity and unbalance of short-text topics and attenuate the effect of occasional pairwise occurrences of words, allowing the model to focus more on the discovery of scarce topics. Unlike previous approaches, CWIBTD uses co-occurrence word networks to model the topic distribution of each word, which improves the semantic density of the data space and ensures its sensitivity in identifying rare topics by improving the way node activity is calculated and normalizing scarce topics and large topics to some extent. In addition, using the same Gibbs sampling as LDA makes CWIBTD easy to be extended to various application scenarios. Extensive experimental validation in the unbalanced short text dataset confirms the superiority of CWIBTD over the baseline approach in discovering rare topics. Our model can be used for early and accurate discovery of emerging topics or unexpected events on social platforms.

## 1   Introduction

Subsequent paragraphs, however, are indented. Topic models are statistical tools for discovering hidden semantic structures in collections of documents [1]. Thematic models and their extensions have been applied in many fields, such as marketing, sociology, political science, and so on [2].

Most topic models are an improvement on the latent Dirichlet Allocation (LDA)[1]. LDA is a hierarchical parametric Bayesian method that can be used to automatically classify or cluster texts into different topics in large corpora. For example, classifying news information according to science and politics, many existing topic models are based on LDA. Specifically, LDA regards documents as a mixture of topics, different words are biased to different topics, and the topics to which texts belong are probability distributions of text words. Statistical inference is then used to learn the probability distribution of the topic to which the word belongs, as well as the probability distribution of the topic to which each document belongs. Generally, LDA-like models group semantically related words into a single topic by using document-level word co-occurrence information[3]. The length and quantity of documents have a huge impact on co-occurrence information, so the length and quantity of documents also have a great impact on LDA. Because the short text contains a low number of words, these models will not be able to get information about how words are related to each other.

With the rapid development of the Internet and various web applications, it is more convenient for people to exchange information, Web search snippets, Weibo, forum messages, news headlines and other short texts have become the main contents of the Internet[4][5]. Specifically, Twitter's roughly 250 million active users generate nearly 500 million tweets a day, These tweets often carry a lot of information about the real world. Therefore, it is very valuable to accurately explore the themes behind these short tweets, including topic detection[6], query suggestions, user interest monitoring[7], document classification, comment summary[8] and text clustering. However, in short text, the context information is severely sparse and word co-occurrence information is lacking, which leads to the performance of traditional topic model in short text. Therefore, short text topic modeling has attracted the attention of machine learning research community in recent years, aiming to overcome the sparsity problem in short text.

In fact, Many people want the LDA-like model to behave better in short texts. For example, related short texts can be aggregated into lengthy pseudo-documents and then inferred about the topic[7], or models trained from external data (such as Wikipedia) can be used to aid thematic reasoning in short texts[9]. In addition, many LDA operations are also introduced, hoping to get the best effect in the short text[10-15]. In 2013, a BITERM theme model similar to LDA for short texts emerged[16], which can deal with short texts well. The Biterm(BTM) topic model is a special hybrid form of Unigram, where LDA is modeled from a single word and BTM is modeled from a pair of words. Another example is the Double sparse topic model[17], which modifies the LDA to understand the focus topic of each short document and the focus term of each topic. In recent years, there have also been some methods combining neural networks to solve short text topic modeling[18].

In short text topic modeling, there are also large application scenarios for rare topics in unbalanced data sets. For example, timely discovery of crisis events on social platforms can greatly reduce losses. Timely detection of rare topics also leads to better prediction of future events. This model works on rare topics in unbalanced data sets in short text modeling.

In this paper, mainly for the network to improve the traditional word co-occurrence, improve the word co-occurrence network activity in the way of calculation, and make reasonable pruning, the network can be reduced due to random paired between words in the document on the result of words, make the identification model for the imbalance in the text scarce theme, and reduce the complexity of the model. Then, the word co-occurrence network is represented back to the pseudo document set, and the topic is found from the co-occurrence network by Gibbs sampling in a way similar to LDA solving text topic.

33

## 2  Related Work

Probabilistic Latent Semantic Indexing (PLSA)[19] and LDA have been widely used in the research of text corpus. In particular, LDA is a more complete generation model Because it adds Dirichlet priors on the basis of PLSA, it makes it more extensible. Many improvements have been made to these two models in the past two decades, such as the dynamic theme model, the social theme model, and the author-theme model[20].

In recent years, with the development of the Internet, the data of short texts has increased in a large range, and the research in this area has attracted more and more attention[24][25][26][27]. Most of the early studies focused on using auxiliary information to increase the data density[28][29][30][31]. For example, Hong and Daviso[7] trained topic models on aggregated tweets that shared the same word and found that these models worked better than models trained directly on the original tweet. Sahami and Heilman[3] proposed a short text similarity measurement method based on search fragments. Yan[16] et al. proposed a special form of Ungram hybrid, called biterm topic model, to improve topic modeling of short texts. Lin et al.[17] added the sparse constraints of document-topic distribution and topic-term distribution on the basis of LDA to model the topic of short texts. Feng et al. used some methods combined with neural network to solve the short text topic modeling method[18], and Van et al.[21] used graph convolutional neural network to solve the short text problem.

For topic modeling of unbalanced texts, prior knowledge has been widely used to alleviate the skewed distribution of documents on different topics. Andrzejewski et al.[22] suggested adding "must link" and "cannot link" constraints into the topic model. Chen et al.[23] use general vocabulary knowledge to help find coherent themes. However, there is still a lack of a perfect scheme for the topic modeling of unbalanced text.

## 3  A Rare Topic Discovery Model for Short Texts Based on Co-occurrence word Network

In this part, we introduce A Rare Topic Discovery Model for Short Texts Based on Co-occurrence word Network. This model is constructed by normalizing scarce topics and large topics in unbalanced samples to a certain extent and converting them into co-occurrence word networks to make full use of context information, and using methods in LDA to make topic prediction.

### 3.1  Co-occurrence Word Network

In a word co-occurrence network (which can also be called a word network below if there is no conflict), nodes are words that appear in a corpus, and the connection of two nodes indicates that the words in these two nodes appear at least once in the same context. The context here can be a single document or a fixed-size sliding window. In this model, in order to limit the size of word co-occurrence network, a fixed-size slid-ing window is selected to preserve only the local context of each word. Sliding Windows of 10 or larger size can fully capture topic similarity between words [32].However, the larger the window size, the higher the computational complexity, as shown in Figure 1. Therefore, we chose to set the sliding window size for normal text and short text to 10. For short texts, you can also simply use each short document as a separate sliding window. We recommend using sliding Windows for short text, especially if the average document length is greater than the window size.

To convert a given set of documents to a word network, we first filter out low-frequency words and stop words, then move the sliding window to scan each document. Any two different words that will appear in the same window will be considered simultaneous. The number of simultaneous occurrences of two words adds up to the weight of the corresponding edge between them. Note that a pair of words may be counted multiple times, which we call the word-pair weighting pattern shown in Figure 1. Words that appeared simultaneously in adjacent locations were counted more often than words that appeared simultaneously far apart. For example, the words W2 and W2 are counted twice, while W0 and W1 are counted only once. At this point, we can encourage models to place adjacent words in the same topic, which can greatly benefit the learning of topic coherence because cohesive words are usually semantically strongly related.

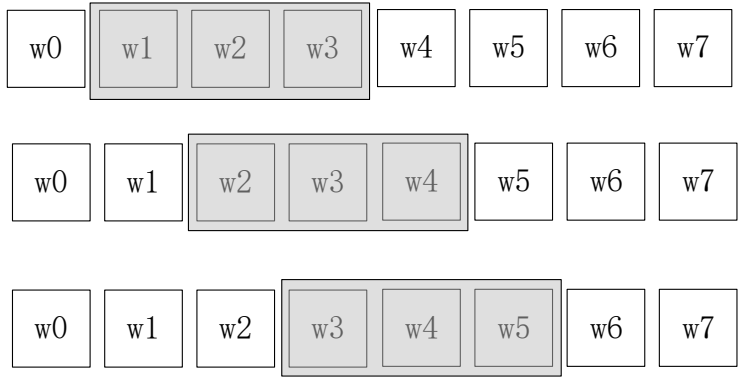

**Fig. 1.** Sliding window process.

Note that in the topic model, a theme can be seen as the words appear frequently in the same document at the same time bags, potential phrases with the words in the network (or community) are very similar, Because words that appear frequently in the same sliding window at the same time are more likely to be related and more likely to belong to the same topic. Therefore, we can use potential phrases in the word network-based model as topics in the LDA. At the same time, it learns the special word space form of topic from word co-occurrence network, and provides theoretical guarantee for topic coherence according to the work of [20].

In order to ensure the model's recognition of unbalanced topic text, PMI scores are used as the activity degree between each two nodes, and the activity degree between nodes $<w_x, w_y>$ is:

$$\text{degree}(w_x, w_y) = \log\left(\frac{p(w_x, w_y)}{p(w_x)p(w_y)}\right) \quad (1)$$

Setting the activity level in this way can eliminate to a certain extent the influence of two words appearing together by chance. Moreover, the influence of word frequency on topic recognition can be weakened for unbalanced text. When the activity is zero, $P(w_x, w_y) = p(w_x) * P(w_y)$, is the occurrence of two words is independent of each other, and the link between the two nodes is cancelled. Positive values indicate that words appear together more often than expected under the independent hypothesis, which means that two words are more likely to belong to the same topic. Negative value indicates that when one word appears, another word is less likely to appear, that is, two words should not belong to the same topic. The node link is cancelled. This method can also prune the original word co-occurrence network and reduce the complexity of the model. The result of this calculation method is shown in Figure 2：

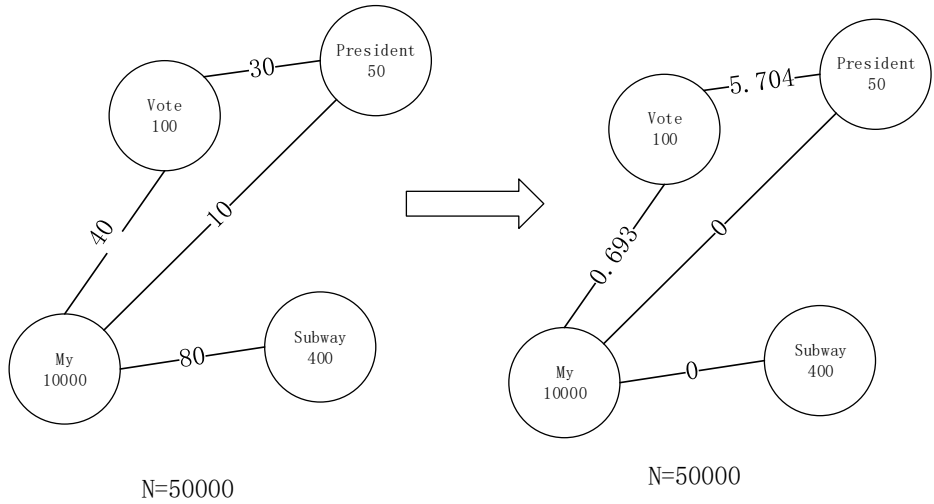

**Fig. 2.** The left figure shows the basic co-occurrence network weight, and the right figure shows the calculation method of co-occurrence network in this model. The calculation method of this model reduces the influence of frequently occurring common words on the model and pays more attention to those words that occur less frequently but more frequently together.

### 3.2 Discover Topics from the Co-occurrence Network

In order to discover topics from co-occurrence networks, we refer to the method in Word Network Topic Model (WNTM)[23][32] and use standard Gibbs sampling to discover potential phrases in large Word networks. We first represent the word co-occurrence network back to the pseudo-document set. Assume that the list of adjacent words for each word is semantically generated by a specific probability model to learn the statistical relationships between words, potential word groups, and lists of adjacent words for words. First, it is assumed that there is a fixed set of potential phrases in the lexical network, and each potential phrase Z is associated with a multinomial distribution on the vocabulary φ _z derived from Dirichlet prior Dir(β). The generation process of the whole pseudo-document set transformed from word network can be explained as follows:

1. For each potential word group Z, the polynomial distribution of Z in the word group $\Phi_z \sim Dir(\beta)$ is obtained

2. The Dirichlet distribution $\theta_i \sim Dir(\alpha)$ of the potential word group of the adjacent word list $L_i$的 of the word $w_i$ is obtained.

3. For each word $w_j \in L_i$:

    - Select a potential phrase $z_j \sim \theta_i$.

    - Select an adjacency $w_j \sim \Phi_{zj}$.

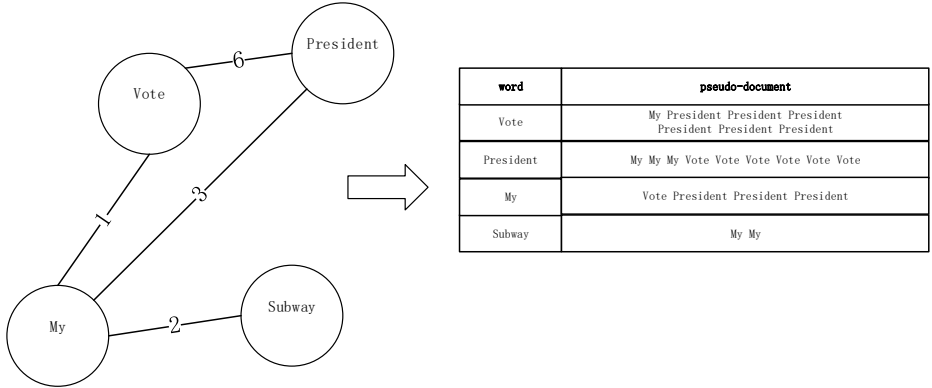

**Fig. 3.** The process of generating virtual text from a co-occurrence network.

In this model, $\theta$ distribution represents the probability that the latent phrase appears in each word's adjacent word list, and Φdistribution represents the probability that the word belongs to each potential phrase. Given the observed corpus, the model first transforms it into a co-occurrence word network, then generates a pseudo-document set, and finally uses the same Gibbs sampling implementation as the traditional LDA to infer the values of potential variables in Φ and $\theta$.

When inferring short text topics, we can use the topic ratio of the word to the adjacent word list $\theta_i$ as the topic ratio in $w_j$. Given the topic ratio for all words, you get the topic for each document. That is:

$$P(z|d) = \sum_{w_i} P(z|w_i)P(w_i|d) \qquad (2)$$

$P(z|w_i) = \theta_{i,z}$, the empirical distribution of the document word as $P(w_i|d)$ estimates, namely:

$$P(w_i|d) = \frac{n_d(w_i)}{Len(d)} \tag{3}$$

Where $n_d(w_i)$ is the word frequency of $w_i$ in document $d$, $Len(d)$ is the length of d.

From the above methods, it can be very straightforward to infer the topic of the passage.

## 4 Experiments

In this section, the datasets and evaluation indicators used in the experiment are introduced.

### 4.1 Datasets

In this section, the datasets and evaluation indicators used in the experiment are introduced.

**Common short text Datasets.** In order to demonstrate the effect of the model, we select the following three datasets [34] to verify the model. After preprocessing these datasets, we present the key information of the datasets summarized in Table 1, where K represents the number of topics in each dataset, N represents the number of documents in each dataset, Len represents the average and maximum length of each document, and V represents the size of the vocabulary.

SearchSnippets: Given eight predefined phrases for different domains, select this data set from the web search transaction results. The eight areas are business, computing, culture and arts, education-science, engineering, health, political-social and sports.

Tweet: At the Text Search Conference (TREC) in 2011 and 2012, 109 queries were available. After deleting queries without highly relevant tweets, the tweet dataset consisted of 89 clusters with a total of 2,472 tweets.

GoogleNews: on the GoogleNews site, news articles are automatically grouped into clusters. The GoolgeNews dataset was downloaded from Google News on November 27, 2013, and captured headlines and snippets of 11,109 news articles belonging to 152 clusters.

Table 1. Basic information about a general data set.

| Dataset        | K   | N      | Len     | V     |
|----------------|-----|--------|---------|-------|
| SearchSnippets | 8   | 12,295 | 14.4/37 | 5,547 |
| Tweet          | 89  | 2,472  | 8.55/20 | 5,096 |
| GoogleNews     | 152 | 11,109 | 6.23/14 | 8,110 |

**Imbalanced short text Datasets.** We process the SearchSnippets dataset by taking the first 300 pieces of data from each of the first four classes and 20 pieces from each of the last four classes to produce an Imbalanced dataset. In this sample, the discovery performance of the model for small categories is verified in the case of large differences in the amount of different types of data. Where K represents the number of topics in each dataset, N represents the number of documents in each dataset, Large represents the number of data and topics contained in each topic of large topic, and rare represents the number of data and topics contained in each topic of rare topic.

**Table 2.** Basic information about the Imbalanced datasets.

| Dataset | K | N | large | rare |
|---|---|---|---|---|
| Imbalance_SearchSnippets | 8 | 1280 | 300/4 | 120/4 |

### 4.2 Evaluation Metrics

By selecting the maximum probability for each document topic, we can get a cluster label for each text. Then, Purity and NMI [22][35] can be used as measurement indicators to compare cluster labels and real labels.

**Normalized mutual information(NMI).** Normalized mutual information (NMI) is used to evaluate the quality of clustering solutions. NMI is an external cluster verification indicator, which can effectively measure the statistical information shared by the random variables representing the cluster assignment and the category assignment of the data points marked by the user. In general, the estimation method of NMI is as follows:

$$NMI = \frac{\sum_{h,l} d_{hl} log(\frac{D \cdot d_{hl}}{d_h c_l})}{\sqrt{(\sum_h d_h log(\frac{d_h}{D}))(\sum_l c_l log(\frac{c_l}{D}))}} \quad (4)$$

Where, $D$ is the number of documents, $d_h$ is the number of documents in class $h$, $c_l$ is the number of documents in group $l$, $d_{hl}$ is the number of documents in class $h$ and group $l$. The NMI value is 1 when the clustering solution exactly matches the user-flagged category assignment, and is close to 0 for random document partitioning.

**Clustering purity.** The main idea of Purity of clustering is to divide the number of correct samples in clustering by the total number of samples. The real category corresponding to each cluster is not known for the result after clustering, so the maximum value in each case is taken. The calculation formula of purity is defined as follows:

$$P = (\Omega, C) = \frac{1}{N} \sum_k max|\omega_k \cap c_j| \quad (5)$$

Where $N$ represents the total number of samples; $\Omega = \{\omega_1, \omega_2, \cdots, \omega_K\}$ said each cluster after clustering, and $C = \{c_1, c_2, \cdots c_j\}$ said the correct category; $\omega_k$ represents all samples in the $k$th cluster after clustering, and $c_j$ represents the real samples in the $j$th category. Here, the value range of $P$ is [0,1]. The larger the value, the better the clustering effect is.

## 5 . Experiments

LDA and WNTM were used as comparison models to test the performance of this model. Because LDA is the basic model in the topic model and WNTM is the original model of this model, this model improves its recognition performance of scarcity categories in unbalanced data sets on the basis of WNTM model. The results of this experiment are averaged after 10 runs.

### 5.1 Parameter Setting

Gibbs Sampling adopted by the model used in this experiment is set to iteration 2000 times, and other super-parameters are set as follows:

LDA: Through grid search, the hyperparameters of LDA are α = 0.05 and β = 0.01[36] to obtain the best performance.

WNTM: For this model, use the hyperparameters α = 0.1, β = 0.1 from the original paper and set the window size to 10 words for best performance.

CWIBTD: For this model, use α = 0.1, β = 0.1 and set the window size to 10 words to get better performance.

### 5.2 Results analysis of General Datasets

Through experiments, the results are shown in the following table:

Table 3. Model performance of each model in general datasets.

| Model | | Google News | Search Snippets | Tweet | Mean Value |
|---|---|---|---|---|---|
| LDA[1] | Purity | 0.793 | 0.740 | 0.821 | 0.785 |
| | NMI | 0.825 | 0.517 | 0.805 | 0.716 |
| WNTM[32] | Purity | 0.837 | 0.712 | 0.856 | 0.802 |
| | NMI | 0.876 | 0.464 | 0.850 | 0.730 |
| CWIBTD | Purity | 0.721 | 0.622 | 0.733 | 0.630 |
| | NMI | 0.734 | 0.452 | 0.796 | 0.541 |

It can be seen from the experimental results that the performance of the CWIBTD model in normal data sets is lower than that of the benchmark model. Through analy-sis of the model, the problem is the cause of our model to increase attention to the scarce topics, themes and scarce for large theme in one operation, similar to the normalized model is missing some information for big subject, so this model for big topic clustering ability decreases. However, in the calculation of the overall clustering evaluation index, large topics occupy too much weight, resulting in a decrease in overall performance. However, the identification performance of this model for scarce topics will be shown in the next section.

### 5.3 Results analysis of Imbalanced Dataset

By experimenting with Imbalance_Search Snippets in an unbalanced dataset, we get the following results: Imbalance_Search Snippets_rare shows how each model behaves in a scarce topic, and Imbalance_Search Snippets_all shows how each model behaves in all topics in the dataset. Through experiments, the results are shown in the following table:

**Table 4.** Model performance of each model in Imbalanced dataset.

| Model | | Imbalance_Search Snippets_rare | Imbalance_Search Snippets_all |
|---|---|---|---|
| LDA[1] | Purity | 0.75 | 0.703 |
| | NMI | 0.645 | 0.434 |
| WNTM[32] | Purity | 0.725 | 0.710 |
| | NMI | 0.635 | 0.471 |
| CWIBTD | Purity | 0.963 | 0.686 |
| | NMI | 0.890 | 0.431 |

It can be seen from the experimental results that the performance of this model is still poor in all topics of unbalanced data sets, but the performance of this model is better improved in scarce topics. It means that this model performs well in the scenario where scarcity theme is found.

## 6 . Conclusion

CWIBTD uses the co-occurrence word network to model the topic distribution of each word, which improves the semantic density of the data space. By improving the calculation method of node activities, CWIBTD normalizes the scarce topic and large topic to some extent, and ensures the sensitivity of the data space to identify the rare topic. However, due to the loss of some data of large topics, the overall performance of the model is low, but the model still has good performance on rare topics. Our model can be used for early and accurate discovery of emerging topics or unexpected events on social platforms.

# References

[1] Blei DM, Ng A, Jordan MI (2003) Latent dirichlet allocation. The Journal of Machine Learning Research
[2] Boyd-Graber J, Hu Y, Mimno D (2017) Applications of Topic Models.
[3] Wang X (2006) Topics over time : a non-Markov continuous-time model of topical trends. In: Proceedings of the 12th ACM SIGKDD International Conference on Knowledge Discovery and Data Mining (KDD '06).
[4] Kou F, Du J, He Y, Ye L (2016) Social network search based on semantic analysis and learning. CAAI transactions on intelligence technology, 1(4):293-302.
[5] Li A, Du J, Kou F, Xue Z, Xu X, Xu M, Jiang Y (2022) Scientific and Technological Information Oriented Semantics-adversarial and Media-adversarial Cross-media Retrieval. arXiv preprint arXiv:2203.08615.
[6] Wang X, Zhai C, Hu X, Sproat R (2007) Mining correlated bursty topic patterns from coordinated text streams. In: Proceedings of the 13th ACM SIGKDD International Conference on Knowledge Discovery and Data Mining, San Jose, California, USA, August 12-15, 2007.
[7] Weng J, Lim EP, Jing J, He Q (2010) Twitterrank: Finding Topic-Sensitive Influential Twitterers. In: Proceedings of the Third International Conference on Web Search and Web Data Mining, WSDM 2010, New York, NY, USA, February 4-6, 2010.
[8] Ma Z, Sun A, Quan Y, Gao C (2012) Topic-driven reader comments summarization. ACM
[9] Xuan HP, Nguyen ML, Horiguchi S (2008) Learning to Classify Short and Sparse Text & Web with Hidden Topics from Large-Scale Data Collections. In: Proceedings of the 17th International Conference on World Wide Web, WWW 2008, Beijing, China, April 21-25, 2008.
[10] Chen Y, Amiri H, Li Z, Chua TS (2013) Emerging topic detection for organizations from microblogs. In: International Acm Sigir Conference on Research & Development in Information Retrieval. p 43
[11] Chua F, Asur S (2013) Automatic Summarization of Events from Social Media. ecological indicators
[12] Zhao W, Jing J, Weng J, He J, Li X (2011) Comparing Twitter and Traditional Media Using Topic Models. In: European Conference on Advances in Information Retrieval.
[13] Li W, Jia Y, Du J (2017) Distributed extended Kalman filter with nonlinear consensus estimate. Journal of the Franklin Institute, 354(17):7983-95.
[14] Li W, Jia Y, Du J (2017) Distributed consensus extended Kalman filter: a variance-constrained approach. IET Control Theory & Applications, 11(3):382-9.
[15] Xu L, Du J, Li Q (2013) Image fusion based on nonsubsampled contourlet transform and saliency-motivated pulse coupled neural networks. Mathematical Problems in Engineering.
[16] Yan X, Guo J, Lan Y, Cheng X (2013) A biterm topic model for short texts. In: International Conference on World Wide Web. pp 1445-1456
[17] Lin T, Tian W, Mei Q, Cheng H (2014) The dual-sparse topic model: mining focused topics and focused terms in short text. In: The Web Conference.
[18] Feng J, Zhang Z, Ding C, Rao Y, Xie H (2020) Context Reinforced Neural Topic Modeling over Short Texts.
[19] Hofmann T (1999) Probabilistic latent semantic indexing. Sigir
[20] (2015) The Author-Topic-Community model for author interest profiling and community discovery. Knowledge & Information Systems 44(2):359-383
[21] Van LN, Xuan BT, ThanKhoat (2022) A graph convolutional topic model for short and noisy text streams. Neurocomputing 468:345-359
[22] Andrzejewski D, Zhu X, Craven M (2009) Incorporating Domain Knowledge into Topic Modeling via Dirichlet Forest Priors. In: International Conference on Machine Learning.
[23] Allahyari M, Kochut K (2016) Discovering Coherent Topics with Entity Topic Models. In:

IEEE/WIC/ACM International Conference on Web Intelligence. pp 26-33.
[24] Wu X, Li C, Zhu Y, Miao Y. (2020) Short text topic modeling with topic distribution quantization and negative sampling decoder. In: Proceedings of the 2020 Conference on Empirical Methods in Natural Language Processing, pp 1772-1782.
[25] Meng D, Jia Y, Du J (2016) Consensus seeking via iterative learning for multi-agent systems with switching topologies and communication time-delays. International Journal of Robust and Nonlinear Control, 26(17):3772-90.
[26] Lin P, Jia Y, Du J, Yu F (2009) Average consensus for networks of continuous-time agents with delayed information and jointly-connected topologies. In 2009 American Control Conference, pp. 3884-3889.
[27] Li Q, Du J, Song F, Wang C, Liu H, Lu C (2013) Region-based multi-focus image fusion using the local spatial frequency. In 25th Chinese control and decision conference (CCDC), pp. 3792-3796.
[28] Shi L, Du J, Liang M, Kou F (2019) Dynamic topic modeling via self-aggregation for short text streams. Peer-to-peer Networking and applications, 12(5):1403-17.
[29] Kou F, Du J, Lin Z, Liang M, Li H, Shi L, Yang C (2018) A semantic modeling method for social network short text based on spatial and temporal characteristics. Journal of computational science, 28:281-93.
[30] Kou F, Du J, Yang C, Shi Y, Liang M, Xue Z, Li H (2019) A multi-feature probabilistic graphical model for social network semantic search. Neurocomputing, 336:67-78.
[31] Shi L, Du J, Liang M, Kou F (2019) SRTM: A Sparse RNN-Topic Model for Discovering Bursty Topics in Big Data of Social Networks. Journal of Information Science & Engineering, 35(4).
[32] Zuo Y, Zhao J, Xu K (2016) Word Network Topic Model: A Simple but General Solution for Short and Imbalanced Texts. Knowledge & Information Systems 48(2):379-398
[33] Arora S, Ge R, Halpern Y, Mimno D, Moitra A, Sontag D, Wu Y, Zhu M (2012) A Practical Algorithm for Topic Modeling with Provable Guarantees. In: International Conference on Machine Learning.
[34] Qiang J, Qian Z, Li Y, Yuan Y, Wu X (2020) Short Text Topic Modeling Techniques, Applications, and Performance: A Survey. IEEE Transactions on Knowledge and Data Engineering(01)
[35] Huang R, Yu G, Wang Z, Zhang J (2013) Dirichlet Process Mixture Model for Document Clustering with Feature Partition. IEEE Transactions on Knowledge and Data Engineering 99(8):1748-1759
[36] Qiang J, Li Y, Yuan Y, Liu W, Wu X (2018) STTM: A Tool for Short Text Topic Modeling.